\documentclass[twocolumn]{aastex631}
\shortauthors{McGloughlin, Steltner, Martins et al.}
\usepackage{ulem}
\usepackage{cellspace}
\usepackage{amssymb}
\usepackage{amsmath}
\usepackage{mathtools}
\usepackage{bm}
\usepackage{longtable}
\usepackage{tabu}
\usepackage{color}
\usepackage{etoolbox}
\usepackage{supertabular}
\usepackage{savesym}
\savesymbol{tablenum}
\usepackage{siunitx}
\restoresymbol{SIX}{tablenum}
\usepackage{ wasysym }
\usepackage{comment}
\usepackage{multirow,makecell}

\newcommand{\Tcoh}{T_{\textrm{coh}}}

\newcommand{\depthbracketrefpop}{45, 65}

\newcommand{\paramtotaltemplates}{\num{1.4e+18}}
\newcommand{\paramtotalWUsmillions}{\num{19.2}}
\newcommand{\nCandPerWUtwoHz}{\num{250000}}
\newcommand{\nCandPerWUfiftymHz}{\num{6250}}
\newcommand{\paramNCandTotal}{\num{4.8e+12}}
\newcommand{\paramWUtotaltemplates}{\num{7.3e+10}}
\newcommand{\paramfmax}{1686.0}
\newcommand{\paramfmin}{800.0}

\newcommand{\TcohFUZero}{\num{60}}
\newcommand{\dfreqmuHzFUZero}{\num{4}}

\newcommand{\dfdotcgTenfHzFUZero}{1.5}
\newcommand{\mskyFUZero}{\num{0.008}}

\newcommand{\dfreqmuHzFUOne}{\num{2}}

\newcommand{\dfdotcgTenfHzFUOne}{0.9}

\newcommand{\mskyFUOne}{\num{0.002}}
\newcommand{\TcohFUOne}{\num{60}}
\newcommand{\SRfreqmuHzFUOne}{\num{1000}}
\newcommand{\SRfdotFUOne}{\num{1}}
\newcommand{\SRskyFUOne}{\num{5}}

\newcommand{\aTwoFrFUOne}{6.76}
\newcommand{\aBSGLtLrFUOne}{0}

\newcommand{\TcohFUTwo}{\num{120}}
\newcommand{\SRfreqmuHzFUTwo}{\num{80}}
\newcommand{\SRfdotFUTwo}{\num{0.3}}
\newcommand{\SRskyFUTwo}{\num{2}}

\newcommand{\TcohFUThree}{coh.}

\newcommand{\NSegFUZero}{73}
\newcommand{\NSegFUOne}{73}
\newcommand{\NSegFUTwo}{37}
\newcommand{\NSegFUThree}{1}

\newcommand{\meanMismatchFUZero}{\num{56}}
\newcommand{\meanMismatchFUOne}{\num{39}}

\newcommand{\NCandFUOne}{\num{13442787}}
\newcommand{\NCandFUTwo}{\num{831387}}
\newcommand{\NCandFUThree}{\num{831387}}
\newcommand{\NCandFUFour}{\num{2}}

\newcommand{\NStarSTwo}{\num{9e7}}
\newcommand{\NStarSThree}{\num{4e4}}

\newcommand{\NCallSTwo}{\num{4.4e5}}
\newcommand{\NCallSThree}{\num{8e4}}

\newcommand{\mostStringentUL}{\num{1.32e-25}}

\newcommand{\hwinjreftimestageseven}{1246070525.0}
\newcommand{\hwinjoneid}{1}
\newcommand{\hwinjonealpha}{\ensuremath{2{:}29{:}34.5244}}
\newcommand{\hwinjonedelta}{\ensuremath{-29{:}27{:}08.8573}}

\newcommand{\hwinjonefreq}{\num{848.93497905}}
\newcommand{\hwinjonefonedot}{\num{-3e-10}}
\newcommand{\hwinjonedfreq}{\num{1.3e-08}}
\newcommand{\hwinjonedfonedot}{\num{-3.6e-15}}
\newcommand{\hwinjonedsky}{\ensuremath{0{:}00{:}00.0669}}
\newcommand{\hwinjsevenid}{7}
\newcommand{\hwinjsevenalpha}{\ensuremath{14{:}53{:}42.1490}}
\newcommand{\hwinjsevendelta}{\ensuremath{-20{:}27{:}02.2693}}

\newcommand{\hwinjsevenfreq}{\num{1220.425863627}}
\newcommand{\hwinjsevenfonedot}{\num{-1.12e-09}}
\newcommand{\hwinjsevendfreq}{\num{1.8e-09}}
\newcommand{\hwinjsevendfonedot}{\num{-1.2e-14}}
\newcommand{\hwinjsevendsky}{\ensuremath{0{:}00{:}00.9693}}

\newcommand{\dparams}{\ensuremath{\bm{\lambda}}}

\newcommand{\fstat}{\ensuremath{\mathcal{F}}}
\newcommand{\semicohfstat}{\ensuremath{\bar{\mathcal{F}}}}

\NewDocumentCommand{\lratio}{ e{^} s o >{\SplitArgument{1}{|}}m }{%
    \operatorname{\Lambda}
    \IfValueT{#1}{{\!}^{#1}}
    \IfBooleanTF{#2}{
        \expectarg*{\expectvar#4}%
    }{
        \IfNoValueTF{#3}{
            \expectarg{\expectvar#4}%
        }{
            \expectarg[#3]{\expectvar#4}%
        }%
    }%
}
\NewDocumentCommand{\lik}{ e{^} s o >{\SplitArgument{1}{|}}m }{%
    \operatorname{\mathcal{L}}
    \IfValueT{#1}{{\!}^{#1}}
    \IfBooleanTF{#2}{
        \expectarg*{\expectvar#4}%
    }{
        \IfNoValueTF{#3}{
            \expectarg{\expectvar#4}%
        }{
            \expectarg[#3]{\expectvar#4}%
        }%
    }%
}
\NewDocumentCommand{\prob}{ e{^} s o >{\SplitArgument{1}{|}}m }{%
    \operatorname{p}
    \IfValueT{#1}{{\!}^{#1}}
    \IfBooleanTF{#2}{
        \expectarg*{\expectvar#4}%
    }{
        \IfNoValueTF{#3}{
            \expectarg{\expectvar#4}%
        }{
            \expectarg[#3]{\expectvar#4}%
        }%
    }%
}
\NewDocumentCommand{\fstatistic}{ e{^} s o >{\SplitArgument{1}{|}}m }{%
    \operatorname{\mathcal{F}}
    \IfValueT{#1}{{\!}^{#1}}
    \IfBooleanTF{#2}{
        \expectarg*{\expectvar#4}%
    }{
        \IfNoValueTF{#3}{
            \expectarg{\expectvar#4}%
        }{
            \expectarg[#3]{\expectvar#4}%
        }%
    }%
}
\NewDocumentCommand{\semcohfstatistic}{ e{^} s o >{\SplitArgument{1}{|}}m }{%
    N_\mathrm{seg} \semicohfstat
    \IfValueT{#1}{{\!}^{#1}}
    \IfBooleanTF{#2}{
        \expectarg*{\expectvar#4}%
    }{
        \IfNoValueTF{#3}{
            \expectarg{\expectvar#4}%
        }{
            \expectarg[#3]{\expectvar#4}%
        }%
    }%
}
\NewDocumentCommand{\expectvar}{mm}{%
    #1\IfValueT{#2}{\nonscript\;\delimsize\vert\nonscript\;#2}%
}
\DeclarePairedDelimiterX{\expectarg}[1]{(}{)}{#1}

\DeclareSIUnit \parsec {pc}

\newcommand{\Gauss}{\mathrm{\MakeUppercase{G}}}
\newcommand{\Signal}{{\mathrm{\MakeUppercase{S}}}}
\newcommand{\Line}{{\mathrm{\MakeUppercase{L}}}}
\newcommand{\Transient}{{\mathrm{t\MakeUppercase{L}}}}

\newcommand{\NoisetL}{{\Gauss\Line\Transient}}

\newcommand{\BSNtsc}{{\hat\beta}_{{\Signal/\NoisetL}}}	

\newcommand{\bsgltl}{\BSNtsc}

\newcommand{\fiftyMHzband}{\SI{50}{\milli\hertz}}

\begin{document}

\title{High-frequency continuous gravitational waves searched in LIGO O3 public data with Einstein@Home}

\correspondingauthor{B. McGloughlin}
\email{brian.mcgloughlin@aei.mpg.de}
\correspondingauthor{J. Martins}
\email{jasper.martins@aei.mpg.de}

\author[0009-0002-4068-7911]{B. McGloughlin}
\author[0000-0003-1833-5493]{B. Steltner}
\author[0009-0002-3912-189X]{J. Martins}
\author[0000-0002-1007-5298]{M. A. Papa}
\author[0000-0001-5296-7035]{H.-B. Eggenstein}
\author[0000-0002-2150-3235]{J. Ming}
\author{B. Machenschalk}
\author[0000-0002-3789-6424]{R. Prix}
\author{M. Bensch}
\affiliation{Max Planck Institute for Gravitational Physics (Albert Einstein Institute), Callinstrasse 38, 30167 Hannover, Germany}
\affiliation{Leibniz Universität Hannover, D-30167 Hannover, Germany}

\begin{abstract}
	We search for nearly-monochromatic gravitational wave signals with frequencies $\SI{\paramfmin}{\hertz} \leq f \leq \SI{\paramfmax}{\hertz}$ \text{and spin-down} $\qty{-2.7e-9}{\hertz\per\second} \leq \dot{f} \leq \qty{0.2e-9}{\hertz\per\second}$. We use LIGO O3 public data from the Hanford and Livingston detectors and deploy this search on the Einstein@Home volunteer-computing project. This is the most sensitive search carried out to date in this parameter space. Our results are consistent with a non-detection. We set upper limits on the gravitational wave amplitude $h_{0}$ and translate these to upper limits on neutron star ellipticity and on r-mode amplitude. The most stringent upper limits are at $\qty{800}{\hertz}$ with $h_{0} = \mostStringentUL$, at the 90\% confidence level. Searching in the high frequency bands allows us to probe astrophysically interesting ellipticities with our results excluding isolated neutron stars rotating faster than $\qty{2.5}{\milli\second}$ with ellipticities $\epsilon \geq$  $1.96 \times 10^{-8}\left[\frac{d}{\qty{100}{\parsec}}\right]$ within a distance $d$ from Earth. Our results also exclude r-mode amplitudes $\alpha \geq 7 \times 10^{-7}\left[\frac{d}{\qty{100}{\parsec}}\right]$ for neutron stars stars spinning faster than 400 Hz.
\end{abstract}

\section{Introduction}
Continuous gravitational waves are expected to be produced by a variety of astrophysical scenarios; from non-axis symmetric rotating neutron stars, destabilization of r-modes \citep{Owen:1998xg}, or from more exotic mechanisms such as emission from axion-like particles surrounding back holes \citep{Zhu:2020tht}.
The expected gravitational wave amplitude at Earth is several orders of magnitude smaller than that of previously detected signals from compact binary coalescence events, but because the signal is long-lasting ($\sim$ years) one can integrate the detector data over many months and increase the signal-to-noise ratio significantly.

The most challenging searches for this type of signal are the all-sky surveys, where one looks for a signal from a source that is not known. In these searches the number of waveforms that can be resolved over months of observation time is very large and so the sensitivity of the search is limited by its computational cost.

Our earlier all-sky searches for continuous gravitational waves from isolated neutron stars have investigated the frequency range below 800 Hz \citep{Steltner:2023cfk}, with a deep search especially devoted to the high sensitivity range of the instruments, between 30-250 Hz \citep{McGloughlin:2025iyx}. In this paper we expand the search space to significantly higher frequencies: $\SI{\paramfmin}{\hertz} \leq f \leq \SI{\paramfmax}{\hertz}$. The resolution in the sky increases with frequency, making this range very hard to cover with high sensitivity.
Thanks to the computing power donated by the volunteers of the Einstein@Home project we are able to match the sensitivity of \citep{Steltner:2023cfk} and with this search ``smoothly" extend its reach well above the 1400 Hz region, where we expect the signal from the fastest known pulsar \citep{Hessels:2006ze}.

The spin-down range of this search -- $\qty{-2.7e-9}{\hertz\per\second} \leq \dot{f} \leq \qty{0.2e-9}{\hertz\per\second}$  -- is higher than \citep{Dergachev:2025hwp} but lower than that scoped in \citep{KAGRA:2022dwb}. This strikes a balance between wanting a high spin-down as high intrinsic spin-down rates can potentially channel more of their lost rotational energy into gravitational wave radiation, and the fact that pulsars typically exhibit spin-downs orders of magnitude smaller than $10^{-9}$ Hz/s.

Finally, the gravitational wave amplitude scales linearly with the degree of deformation of the source, the ellipticity, so that at high frequency less deformed sources are detectable comparable with lower frequencies.

We employ a hybrid follow-up procedure on our signal candidates in which we combine a traditional frequentist grid-based method, used for instance in \citet{Steltner:2023cfk}, followed by a  Bayesian non-deterministic follow-up \citep{Martins:2025jnq}. Our results are consistent with a non-detection and we set upper limits on the intrinsic gravitational wave amplitude accordingly.

This paper is organized as follows: In Section \ref{sec:Signal}
we describe the signal model, in Section \ref{sec:Generalities} we outline the generalities of this search, in Section \ref{sec:EAHSearch} we describe the initial stage of the search run on Einstein@Home, in Section \ref{sec:FollowUp Searches} we describe the hybrid follow-up procedure of interesting candidates resulting from the initial search, in Section \ref{sec:Results} we present our results, and finally in Section \ref{sec:Conclusions} we discuss the astrophysical implications of our search results.

\section{The signal}\label{sec:Signal}

This search targets quasi-monochromatic gravitational wave signals described in \citep{Jaranowski:1998qm,Cutler:2005hc}. These signals display both frequency and amplitude modulations due to the motion of the Earth. The waveforms take the following form:
\begin{equation}
	h(t) = F_{+}(\alpha, \delta, \psi;t)h_{+}(t) + F_{\times}(\alpha, \delta, \psi;t)h_{\times}(t),
\end{equation}
where $F_{+}(\alpha, \delta, \psi;t)$ and $F_{\times}(\alpha, \delta, \psi;t)$ are called the detector beam-pattern functions for both the ``$+$" and ``$\times$" wave polarizations and $h_+$ and $h_\times$ are the polarization amplitudes.
$F_+$ and $F_\times$ depend on the inclination and declination of the source $(\alpha,\delta)$, the polarization angle $\psi$ as well as t, the time at the detector.
The plus and cross waveforms are
\begin{eqnarray}
	h_{+}(t) = A_{+}\cos\Phi(t) \nonumber\\
	h_{\times}(t) = A_{\times}\sin\Phi(t)
\end{eqnarray}
with amplitudes
\begin{eqnarray}
	A_{+} &=& \frac{1}{2}h_{0}(1 + \cos^{2}{\iota}) \nonumber\\
	A_{\times} &=& h_{0}\cos{\iota}.
\end{eqnarray}
Here, $h_{0} \ge 0$ is the intrinsic gravitational wave amplitude, $0 \le \iota \le \pi$ is the angle between the angular momentum vector of the neutron star and the line of sight and $\Phi (t)$ is the phase of the gravitational wave at time t.

The phase evolution can be expressed in the solar-system barycenter (SSB) frame as:
\begin{equation}
	\Phi(\tau_{\textrm{SSB}}) = \Phi_{0} + \newline\\ 2\pi[f(\tau_{\textrm{SSB}} - \tau_{0SSB}) + \frac{1}{2}\dot{f}(\tau_{\textrm{SSB}} - \tau_{0,\textrm{SSB}})^{2}].
\end{equation}
The reference time for the phase evolution is taken as $\tau_{0, \textrm{SSB}} = 1246070525.0$.

\section{Search Generalities}
\label{sec:Generalities}

\subsection{The Data}
This search uses LIGO public data from the first half of the third observation run (O3a), i.e. between GPS time 1238166018 (Apr 01 15:00:00 GMT 2019) and 1254150018 (Oct 03 15:00:00 GMT 2019).

Following \citet{Steltner:2021qjy} and previous Einstein@Home searches, we remove noise that would degrade the quality of the search results, lines and in the frequency domain and loud glitches in the time-domain.

\begin{figure}[h!tbp]
	\includegraphics[width=\columnwidth]{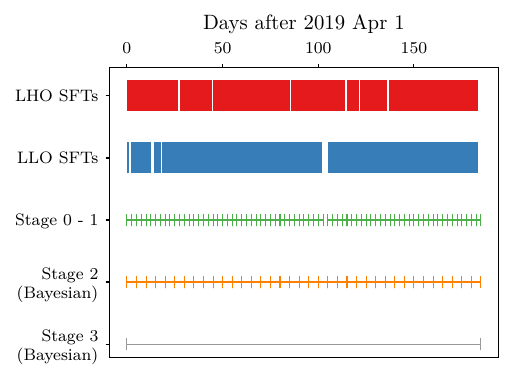}
	\caption{Segmentation for the semi-coherent grid based searches of Stages 0-1, Stages 2-3 Bayesian searches and the input data (SFTs in LHO and LLO). Stage 3 is a fully-coherent search on O3a data.}
	\label{fig:NsegSegmentation}
\end{figure}

The input to our searches is in the form of half-hour time-baseline Fourier Transforms (SFTs) as shown in Figure \ref{fig:NsegSegmentation}.

\subsection{Hierarchical Semi-coherent Searches}

All-sky searches for continuous gravitational wave signals over a broad frequency range are computationally unfeasible using optimal detection methods. Semi-coherent hierarchical schemes have been suggested in order to achieve the best sensitivities under computational budget constraints \citep{Brady:1997ji,Brady:1998nj,Krishnan:2004sv} and are nowadays commonly used \citep{KAGRA:2022dwb,McGloughlin:2025iyx,Dergachev:2025ead}, to cite recent examples.

A semi-coherent approach entails coherent searches over a baseline $\Tcoh$, shorter compared to the entire observation span, and then the incoherent combination of the results from the various $\Tcoh$ segments, typically performed by summing the values of the coherent detection statistic. We use the semi-coherent Global Correlation Transform (GCT) search proposed by \cite{Pletsch:2009uu,Pletsch:2010xb}, with the multi-detector $\fstat$-statistic of \citep{Cutler:2005hc} as the per-segment coherent detection statistic:
\begin{equation}
	\label{eq:semicohFstat}
	\semicohfstat(\dparams) := \frac{1}{N_{seg}}\sum_{i=1}^{N_{\text{seg}}}\mathcal{F}_{i}(\dparams_i).
\end{equation}
The parameter space points $\dparams_i$ that enter in this summation \eqref{eq:semicohFstat} depend on the template parameters $\dparams$ of the left-hand-side of  Eq.~\ref{eq:semicohFstat}. The $\dparams_i$ are on a grid that is coarser than the grid for $\dparams$. This is because the resolution increases with a longer observation baseline. The  GCT is an efficient way to pick the set $\{\dparams_i\}$ to approximate the detection statistic at $\dparams$.

\subsection{Detection Statistics}

For a rotating isolated neutron star the signal waveform is described by the eight parameters defined is Section \ref{sec:Signal}, but the search template waveform only depends on the four phase-evolution parameters $\dparams$: the frequency $f$, the spin-down $\dot f$ and the two sky coordinates of the source $\alpha$ and $\delta$ \citep{Jaranowski:1998qm}.

The $\mathcal{F}$-statistic is the maximum log-likelihood ratio between the signal hypothesis and a pure Gaussian noise hypothesis. Since it is possible to have large $\mathcal{F}$-statistic values due to \textit{non-Gaussian} disturbances of the data, we also utilize the more robust detection statistic $\bsgltl$ \citep{Keitel:2015ova}, which tests the signal hypothesis against an expanded noise hypothesis i.e. “G” Gaussian noise or “L” lines or “tL” transient lines.

In this paper we use a grid-based deterministic approach in the first 2 stages (Stages 0-1) of the hierarchy and a stochastic approach for the last two stages (Stages 2-3). Both approaches use the detection statistic of Eq.~\ref{eq:semicohFstat}. The first stages, where it is crucial to identify disturbance, are also augmented by the use of the $\bsgltl$ statistic.

\begin{deluxetable*}{lcccccccccccccccccc}
	\tablecaption{Overview of the full hierarchy of grid-based and Bayesian searches. Shown are the values of the following parameters: the stage number, the coherent time baseline $T_\mathrm{coh}$ of each segment and the number of segments $N_\mathrm{seg}$; the parameter space ranges searched around each candidate, $\Delta f, \Delta \dot{f}$ and ${\textrm{r}_\textrm{sky}}$. The radius  ${\textrm{r}_\textrm{sky}}$ is expressed in units of the side of the hexagon sky-grid tile of the Stage 0 search (Eq.~\ref{eq:skyGridSpacing});
		the number of candidates searched (N$_{\textrm{in}}$) and how many of those survive and make it to the next stage (N$_{\textrm{out}}$). For the grid searches, shown are also the coarse grid spacings $\delta f,\delta \dot{f}$ and $m_{\text{sky}}$ and of the fine grid $\delta \dot{f}_\mathrm{fine}$; and the average mismatch $\left<\mu \right>$.
		For the Bayesian stages, shown are also the prior-distribution types; the number of live-points $n_\mathrm{live}$; the effective parameter space volume searched per candidate $\langle N_\star \rangle$; and the average number of statistic evaluations taken per candidate $\langle N_\mathrm{call}\rangle$.
		\label{tab:FUtable}
	}
	\tablehead{
	\colhead{Search} & \colhead{$T_\mathrm{coh}$} & \colhead{$N_{\mathrm{seg}}$} & \colhead{$\delta f$}         & \colhead{$\delta \dot{f}$}              & \colhead{$\gamma_1$}   & \colhead{$m_{\text{sky}}$}     & \colhead{$\left<\mu \right>$ } & \colhead{$\Delta f$}         & \colhead{$\Delta \dot{f}$}                  & \colhead{$ {\textrm{r}_\textrm{sky}\over {d(\mskyFUZero )}}$} & \colhead{ N$_{\textrm{in}}$ } & \colhead{N$_{\textrm{out}}$} \\
	Grid             & hr                         &                              & \colhead{$\mu{\textrm{Hz}}$} & \colhead{$10^{\scriptstyle{-10}}$ Hz/s} &                        &                                & \colhead{$10^{-2}$}            & \colhead{$\mu{\textrm{Hz}}$} & \colhead{{{$10^{\scriptstyle{-10}}$} Hz/s}} &                                                               &
	}
	\startdata
	Stage 0          & $\TcohFUZero$              & $\NSegFUZero$                & $\dfreqmuHzFUZero$           & $\dfdotcgTenfHzFUZero$                  & $50$                   & $\mskyFUZero$                  & \meanMismatchFUZero            & \tiny{full range}            & \tiny{full range}                           & \tiny{all-sky}                                                & $\paramtotaltemplates$        & $\NCandFUOne$                \\
	Stage 1          & $\TcohFUOne$               & $\NSegFUOne$                 & $\dfreqmuHzFUOne$            & $\dfdotcgTenfHzFUOne$                   & $450$                  & $\mskyFUOne$                   & \meanMismatchFUOne             & $\SRfreqmuHzFUOne$           & $\SRfdotFUOne$                              & $\SRskyFUOne$                                                 & $\NCandFUOne$                 & $\NCandFUTwo  $              \\
	\hline
	\hline
	Bayesian         &                            &                              & \colhead{Prior}              & \colhead{$n_\mathrm{live}$}             & $\left<N_\star\right>$ & $\left<N_\mathrm{call}\right>$ &                                &                              &                                             &                                                                                                                              \\
	\cline{1-1}\cline{4-7}
	Stage 2          & $\TcohFUTwo$               & $\NSegFUTwo$                 & Jeffreys'                    & 750                                     & \NStarSTwo             & \NCallSTwo                     &                                & $\SRfreqmuHzFUTwo$           & $\SRfdotFUTwo$                              & $\SRskyFUTwo$                                                 & $\NCandFUTwo$                 & $\NCandFUThree$              \\
	Stage 3          & \TcohFUThree               & $\NSegFUThree$               & GMM                          & 250                                     & \NStarSThree           & \NCallSThree                   &                                & -                            & -                                           & -                                                             & $\NCandFUThree$               & $\NCandFUFour $              \\
	\enddata
\end{deluxetable*}
\subsubsection{Search Grids}

The Stage 0 and Stage 1 coarse grids in frequency and spin-down are each described by a single parameter, the grid spacing, which is constant over the search range. The sky grid is approximately uniform on the celestial sphere orthogonally projected on the ecliptic plane. The tiling is a hexagonal covering of the unit circle with hexagon edge length
\begin{equation}\label{eq:skyGridSpacing}
	d(m_{\textrm{sky}}) = 0.15 \sqrt{m_{\textrm{sky}}}\left[\frac{100 \textrm{Hz}}{f}\right],
\end{equation}
where $m_{\textrm{sky}}$ is a tunable free parameter that controls the coarseness of the grid.

The fine grid is the same as the coarse grid in frequency and sky, and it is refined in spin-down by a factor $\gamma_1$.

Table~\ref{tab:FUtable} details the set-ups for the entire hierarchy of searches.

\label{subsec:grids}

\section{Stage-0}
\label{sec:EAHSearch}

\subsection{Sensitivity Goal and Search Set-up}
\label{subsec:sensgoal}

Since a signal that is not detected at Stage 0 can never be recovered at later stages, the most consequential choices affecting the sensitivity of the search are those defining its earliest stage, in particular the $\Tcoh$, the grids and the clustering procedure that selects the candidates to follow-up. Ensuring a high detection efficiency at this stage while keeping the computational cost at bay is difficult because the first stage of the search has to investigate the entire parameter space, and it is hence the most computationally challenging of the whole hierarchy.

Our sensitivity goal is to match the sensitivity depth \citep{Behnke:2014tma} of \citep{Steltner:2023cfk}, $\mathcal{D} \approx 56~[{1/\sqrt{\textrm{Hz}}}]$, which is non trivial because we are here probing much higher frequencies. We consider different $\Tcoh$ and grid combinations, and estimate their detection efficiency on a population of target signals.

This target population comprises signals with intrinsic amplitudes $h_0$ drawn from a uniform distribution constrained to the targeted sensitivity depth bracket
\begin{equation}
	\mathcal{D} = \left[\depthbracketrefpop\right]~[{1/\sqrt{\textrm{Hz}}}].
\end{equation}

We estimate the sensitivity of the search in terms of the detection efficiency as a function of sensitivity depth. The detection criterion is that a signal is detected if it survives the first stage cut. This is of course a necessary condition, and since the follow-up is designed to not lose any signal from the target population that has passed the first cut, it is also sufficient.

As described in Section \ref{subsec:Stage0Post}, the first cut is made based on detection statistic values of uncorrelated candidates. The de-correlation happens through clustering, which is hard to model in detail.
We thus implement a simpler and conservative estimate of the detection efficiency.

We draw a signal at a fixed sensitivity depth from our target population and we imagine performing the Stage 0 all-sky search and clustering in a small volume of parameter space around the signal parameters, on data containing only noise, and on data containing noise+signal. Let's say that this search would require $N^{small}$ templates.

Let's indicate with $N^{tot}$ the number of templates covered by the entire search, with $N_{FU}^{tot}$ the total number of candidates that we can follow-up, and their ratio with $k=N_{FU}^{tot}/N^{tot}$. If the entire parameter space volume is divided in small search volumes, then on average each will contribute $k\cdot N^{small}$ follow-up candidates -- the loudest ones. We take the small volume to be such that this expected number is 1, and set as a detection criterion that the loudest from the search with the signal has to be louder than the loudest from that from the noise-only search.

So, to estimate the detection efficiency we draw from the distribution $f_{max,n}(x^\ell)$ of the loudest detection statistic $x^\ell$ over $n=N^{small}$ realisations, in two cases, noise-only and noise+signal. We compare the two, and consider the signal recovered if the signal draw is higher than the noise draw. We do this 2000 times and the detection efficiency is simply the ratio of the number of recovered signals to 2000.

The probability density function for the loudest takes the form
\begin{equation}\label{maxDSpdf}
	f_{max,n}(x^\ell) = nF^{n-1}(x^\ell)f(x^\ell),
\end{equation}
where $x^\ell$ is the maximum over $n$ trials of a random variable with probability density function $f(x)$ and cumulative probability function $F(x)$. In our case the random variable is $N_{seg} \semicohfstat$ of Eq.~\ref{eq:semicohFstat}, so $f(x)= \chi^2_{4N_{seg}}(x,\rho^2)$ is a chi-square distribution with $4N_{seg}$ degrees of freedom and a non-centrality parameter $\rho^2=\rho_{sign}^2 (1-\mu)$. The mismatch value $\mu$ is drawn from the mismatch distribution of the search grid and $\rho_{sign}^2$ is the non-centrality parameter in the case when there is no mismatch between signal and template grid point.

This estimate of the detection efficiency is conservative for two reasons: 1) the loudest is taken on $N^{small}$ templates that are not independent. The first stage cuts are instead applied to de-correlated, hence independent, candidates. This means that in order to be detected in a real search the signal has to be louder than the loudest in a smaller set of trials than the ones simulated in our procedure, hence smaller signals are actually detectable. 2) the clustering does more than just de-correlating \citep{Steltner:2022aze} and so the signal candidates selected by clustering are often below the ``loudest" level of our simulation (see figure 5 of \citep{Steltner:2023cfk}). These effects result in an underestimate of the detection efficiency by about 15\%.

\begin{figure}[h!tbp]
	\includegraphics[width=\columnwidth]{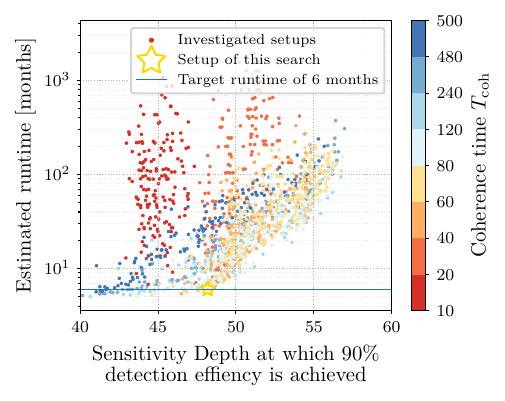}
	\caption{The x-axis shows the amplitude of signals detectable by each Stage 0 search set-up with 90\% confidence. The y-axis is the estimated cost for the search reported for that Stage 0 search set-up. The horizontal line indicates our computational budget. We choose the most sensitive search set-up within our budget (the star).}
	\label{fig:SensDepthVsRuntime}
\end{figure}

Figure \ref{fig:SensDepthVsRuntime} shows the run-time versus sensitivity depth achieved by different set-ups at 90\% detection efficiency. The chosen search set-up has an estimated sensitivity depth of $\approx 48~[{1/\sqrt{\textrm{Hz}}}]$. After validation by injection-and-recovery studies with clustering applied, the actual sensitivity depth of our chosen set-up at 90\% detection efficiency is measured to be about 16\% higher, in line with our sensitivity goal of
$56~[{1/\sqrt{\textrm{Hz}}}]$.

\subsection{Computational load Distribution on Einstein@Home and Atlas}

Of the total computational cost of the Stage 0 search, roughly a fifth was carried out on the Atlas computer cluster at the Max Planck Institute for Gravitational Physics, and the remainder on the Einstein@Home project\footnote{\url{https://einsteinathome.org}} which uses the idle time on individual volunteers computers to conduct large-scale computationally intensive searches \citep{Boinc2, Boinc3}.
Both of these search components are run on Graphical Processing Units (GPUs); The \texttt{NVIDIA Geforce RTX 2070 SUPER} model exclusively on ATLAS while a broad and diverse range of GPU models is used on Einstein@Home.

The total number of templates covered by Stage 0 is {\paramtotaltemplates}. The search is divided into work-units (WUs), where each WU searches approximately 150 sky points, $2\textrm{ Hz}$ in frequency and the entire spin-down range, totaling {\paramWUtotaltemplates} templates each. A total of {\paramtotalWUsmillions} million WUs are required to cover the entire parameter range.

Each WU returns the top \nCandPerWUfiftymHz{} results per \fiftyMHzband, so in total \nCandPerWUtwoHz. The reason for using separate ``top-lists" for each \fiftyMHzband {} is so that saturation effects from loud disturbances which produce many high-ranking results are effectively confined.

The average runtime of a WU on an ATLAS GPU is approximately 30 minutes, implying a total runtime cost of 1100 GPU-years. This search ran on Einstein@Home for approximately 9 months.

\begin{figure}[h!tbp]
	\centering
	\includegraphics[width=\columnwidth]{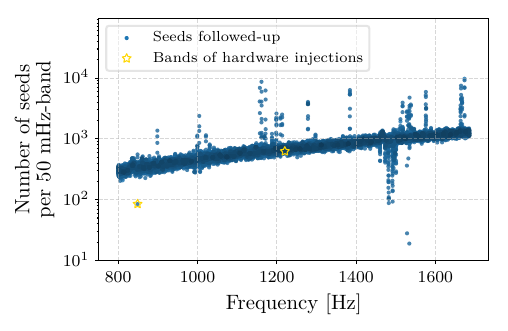}
	\caption{The number of Stage 0 candidates from every \SI{50}{\milli\hertz} band. The outliers are discussed in Section \ref{subsec:Stage0Post}. The relatively large amplitude of hardware injection 1 in the $\SI{848.90}{\hertz}$ band saturates the results and produces fewer but larger clusters in this \SI{50}{\milli\hertz} band.
	}
	\label{fig:seedsPerBand}
\end{figure}

\subsection{Post-processing}
\label{subsec:Stage0Post}

The following steps are performed on the WU-results returned by the Stage 0 search:

\begin{description}
	\item[Banding]
	      All results from the same \SI{50}{\milli\hertz} are collated together. These results cover the full spin-down range and the whole sky in that frequency band.
	\item[Visual Inspection]
	      All \SI{50}{\milli\hertz} results are visually inspected to identify disturbed bands \cite{LIGOScientific:2017wva}. We do not set upper limits in these bands because the detection efficiency is hard to assess but seeds from these bands are followed up, in the spirit of not leaving any stone unturned.
	\item[Clustering]
	      Nearby templates are not completely independent. Thus disturbances or signals tend to contribute several nearby results, whereas statistical fluctuations rarely form clusters. With our clustering technique \citep{Steltner:2022aze} we identify results that are likely due to the same underlying cause, and from there on treat them as a single candidate.

	      A cluster seed is identified, i.e. a set of signal parameters ${\dparams}_{\textrm{seed}}$ and an uncertainty range $\Delta{\dparams}$, such that, if the cluster were due to a signal, with high confidence the signal parameters ${\dparams}_{s}$ would fall in the interval ${\dparams}_{\textrm{seed}}\pm \Delta{\dparams}$.
	      \par\noindent
	      Clustering reduces the $\paramNCandTotal$ results to a more manageable set of order $13$ million seeds.
	      The clustering parameters are determined based on our target sensitivity goal.

\end{description}

The $13$ million seeds are binned in 50 mHz bands and Figure \ref{fig:seedsPerBand} shows how many seeds are found in each band. There exist large excursions from the mean number of seeds, both in excess and in defect. The large positive excursions either come from bands flagged as disturbed by the visual inspection of the results or from bands that house coherent lines which the detection statistic down-weights, but the clustering recuperates as possible weak candidates. The defect in the number of seeds may arise when there are very loud signals or signal-like artefacts that saturate the returned results and conglomerate in few clusters with a large number of occupants. This is the case of the hardware injection 1 at about 850 Hz: its 50 mHz band displays a very low number of seeds compared to most of the other bands. Between 1460-1530 Hz the cause for the small number of seeds is not completely clear, but it is correlated with a steep slope in the power spectral density of the input data which always happens in the surviving shoulders of lines that were cleaned out based on the known lines list \cite{known-lines-list}. We exclude these 50 mHz from the upper limit statement.

\section{Follow-Up}\label{sec:FollowUp Searches}

All follow-up stages are run exclusively on the CPUs on the ATLAS computing cluster.

\subsection{The First Follow-up Stage}
\label{sec:gridFollowUp}

The first follow-up aims at decreasing the uncertainty around each candidate seed. This is achieved with the same coherent time baseline as Stage 0, but a finer grid. From each search the loudest $\bsgltl$ result is taken as the Stage 1 candidate.

\begin{figure}
	\includegraphics[width=\columnwidth]{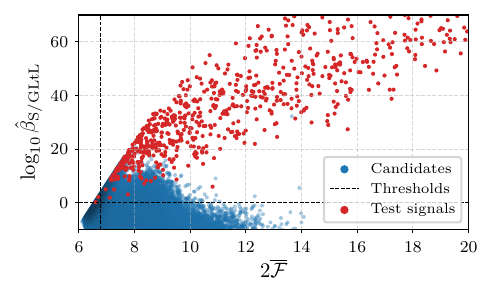}
	\caption{The overlap of the detection statistic distributions of the candidates and the reference signal population at Stage 1. Dashed lines mark the thresholds used to veto candidates.
		The two test signals below the ${2\bar\fstat_r}$ threshold are the two signals we dismiss (see section \ref{sec:gridFollowUp}.)
		For the candidate-distribution, a darker shading implies a higher density of points on a log-scale.}
	\label{fig:gridFollowUp}
\end{figure}

Figure~\ref{fig:gridFollowUp} shows the Stage 1 candidates from the search and from the target signal population.
The distributions for the candidates and the reference population have not yet separated, and further follow-up is necessary.

Based on the signal population results, we define a veto that identifies candidates that should pass on to the next stage: a candidate survives if
\begin{equation}
	\begin{cases}
		2\bar{\mathcal{F}}^{\textrm{cand}} \geq 2\bar{\mathcal{F}}^{\mathrm{thr}} & = \aTwoFrFUOne     \\
		\log_{10}\bsgltl^{\textrm{cand}}\geq \log_{10}\bsgltl^{\mathrm{thr}}      & =  \aBSGLtLrFUOne.
	\end{cases}
\end{equation}

With these thresholds, approximately $94\%$ of the candidates are vetoed, leaving about $800$-thousand candidates to be examined in the next stage.
The thresholds also veto the two dismissed test signals.

We evaluate the uncertainty around each candidate from the signal target population.
Two test signals Stage 1 candidates are offset from their true parameter values significantly more than the other signals.
We set the search region for the next stage ignoring these two signals and accepting
their loss, as their inclusion would incur an unreasonably computational expense in the next stage.

\subsection{The Bayesian Follow-up Method}

Starting from Stage 2, we employ the Bayesian follow-up method of \cite{Martins:2025jnq}.

The uncertainty region $\Delta\dparams$ from Stage 1 is searched at each subsequent Stage with a stochastic method, which produces an estimate of the posterior probability density of the phase-evolution parameters $\dparams \in \dparams_\mathrm{seed} \pm \Delta\dparams$ and of the evidence. The evidence $Z$ measures the support for the presence of a signal.
This posterior is used as the prior for the next stage. At the initial Monte Carlo stage we use uninformative priors following Jeffreys' invariance principle \citep{Jeffreys:1946,Martins:2025jnq}.

We use a nested sampling algorithm \citep{Skilling:2004pqw} -- specifically \texttt{dynesty} \citep{Speagle:2019ivv} wrapped by the \texttt{bilby}-package \citep{Ashton:2018jfp}.
Within the algorithm, we use a \texttt{Bilby}-native implementation of a differential-evolution random-walk (``diff'') to propose new samples.

We use $Z^a$, where $a=2,3$ identifies the stage, as a test statistic to identify false alarms and significant candidates.
Following the description in \textcite{Martins:2025jnq}, we also define a test-statistic $R^{3}$ based on the change of $\log Z^3$ relative to the first Bayesian Stage $2$:
\begin{equation}\label{eq:R}
	R^{3} = \frac{\log Z^{3} - \log Z^{2}}{\log Z^{2}}.
\end{equation}
This statistic has proven effective in identifying noise disturbances, since $Z^a$ remains roughly constant for our target signals whereas it tends to decrease when evaluated on longer time baselines (higher ``a") on search  candidates.

The method recovers all signals of the target population that survived Stage 1, with a $T_{\textrm{coh}} = \TcohFUTwo$ hours at Stage 2 and with a fully coherent search at Stage 3.
The corresponding effective parameter space volumes searched ($N_\star$ as listed in Table~\ref{tab:FUtable}) agree with our findings in \cite{Martins:2025jnq}.

\subsection{Bayesian Follow-up Results}
\label{sec:BayesianFUResults}

\begin{figure}[tb]
	\includegraphics[width=\columnwidth]{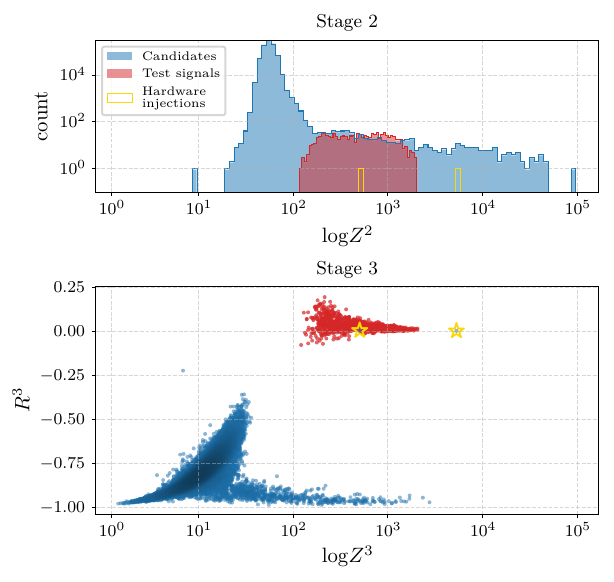}
	\caption{The results of the Bayesian searches in Stage 2 and 3 for the signal candidates and the reference population of test signals. The upper plot shows the evidences at Stage 2, $Z^2$.
		For Stage 3, we show the integrated evidences $Z^3$ and their relative change from Stage 2, $R^3$ (Eq.~\ref{eq:R}), as a scatter plot.
	}
	\label{fig:stochasticFollowUp}
\end{figure}

At Stage 2 the evidences estimated from the reference signal population are not significant compared to the search candidates, which we expect to be noise-dominated (Fig.~\ref{fig:stochasticFollowUp}, top panel).
At Stage 3, conversely, the bulk of the candidate distribution has clearly separated from the reference signal population and has significantly smaller evidences and smaller $R^3$ values (Fig.~\ref{fig:stochasticFollowUp}, bottom panel).

Three features of the candidate distribution strike as deviations from the bulk behaviour: 1) Two candidates are compatible with the test-signal population.
Both candidates correspond to fake signals added to the data at the detector level, the so-called hardware injections, as described in the next section.
2) A long ``tail'' in the candidate $R^3-\log Z^3$ plot, comprising of order $700$ candidates with evidences consistent with the signal reference population, but with $R^3 \approx -1$, inconsistent with the signals. In other words the evidence at Stage 3 is high, but it was much higher at the previous, less sensitive stage. All of these outliers come from only a few frequency bands. An inspection of the input data and of the results from these bands finds that they are plagued by non-Gaussian noise features in one of the two detectors, with many of these being remnants of insufficient line cleaning caused by too-narrow line-cleaning windows.
In most bands, the noise features are isolated and give rise to only a small number of outliers ($\sim$ tens).
However, about 300 candidates arise from the same 1500-1501 Hz band, infested by the violin modes \citep{LIGO:2021ppb}. The visual inspection of the Stage 0 results did not flag this as a disturbed band, and in fact the Stage 1 results from these bands are also largely unremarkable. The first stochastic stage, Stage 2, with much higher effective resolution than the previous stages and not using the line-robust statistic $\bsgltl$, produces significant values of the evidence, which however decrease in the last fully coherent stage. This behaviour is fully consistent with a disturbance. The effect of these disturbances is so pervasive in the 1500-1501 Hz band that we cannot confidently evaluate the detection efficiency and hence do not set upper limits (see Section~\ref{sec:upperlimits}) for them.
3) An outlier with unremarkable evidence $Z^3$, but relatively high $R^3$, albeit smaller than observed for the reference population.
This candidate stands out because of the very low $Z^2$, in fact the top panel of Fig.~\ref{fig:stochasticFollowUp} shows that it has the overall smallest Stage-2 evidence of $\approx 9$. Like for some of the outliers discussed above, this also was caused by insufficiently broad cleaning of a violin mode at $1510.9$ Hz.

These results underline the importance of the identification, characterization, and cleaning of non-Gaussian noise prior to a search \citep{Covas:2018oik,Davis:2018yrz,LIGO:2021ppb,Capote:2024rmo}.

\section{Results}\label{sec:Results}

\subsection{Hardware Injection Recovery}
The hardware injections are fake signals added to the data by using photon calibrators to physically move the detector mirrors in order to provide a check of the entire detection chain \citep{Biwer:2016oyg}. These hardware injections serve as reference signals that provide standard detection benchmarks for any continuous wave search pipeline.

\begin{figure}[ht]
	\includegraphics[width=\columnwidth]{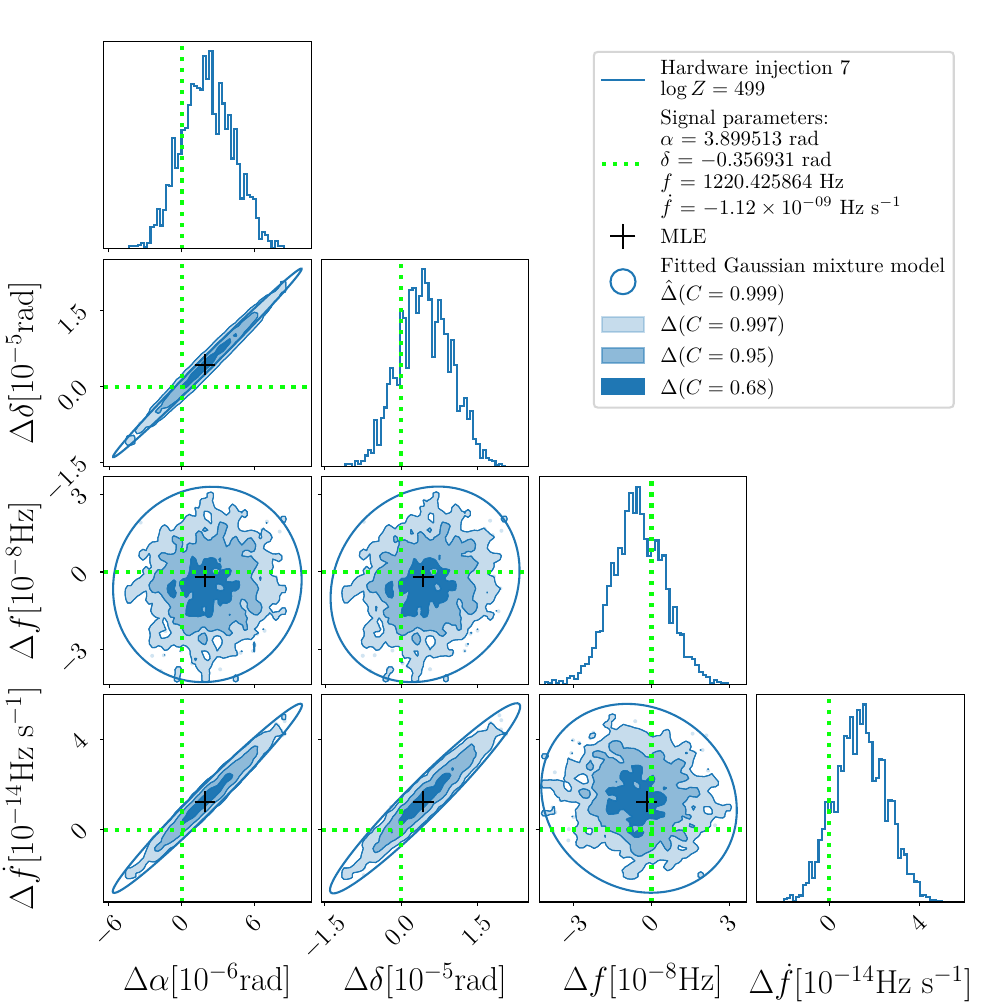}
	\caption{The posterior distribution for hardware injection 7 after search Stage 3. The signal parameters are given at our reference time \hwinjreftimestageseven~(GPS time).}
	\label{fig:hardware_inj_posterior_7}
\end{figure}

Two hardware injection signals from isolated neutron stars fall in our search range, specifically those with IDs 1 and 7 \citep{O3_injection_params}. We recover both, marking the first broad survey to recover hardware injection 7. For this reason, in Figure \ref{fig:hardware_inj_posterior_7}, we show the Stage 3 posteriors for the phase-evolution parameters.

\begin{deluxetable*}{lccccccc}
	\tablecaption{Frequency $f$, frequency derivative $\dot{f}$ and sky position $\alpha, \delta$ of the hardware injections and the offsets between these and the maximum likelihood estimators recovered by our coherent Monte Carlo search using O3a data (Stage $3$). The frequency $f$ is given at the reference time \hwinjreftimestageseven~(GPS time).  \label{tab:HIRecoveryMonteCarlo}}
	\tablehead{
	\colhead{ID$_{\rm inj}$} & \colhead{ $f$}  & \colhead{$\dot{f}$} & \colhead{$\alpha$} & \colhead{$\delta$} & \colhead{$\Delta f$} & \colhead{$\Delta \dot{f}$} & \colhead{Sky distance} \\
	                         & [Hz]            & [Hz/s]              & [hr:m:s]           & [deg:m:s]          & [Hz]                 & [Hz/s]                     & [deg:m:s]
	}
	\startdata
	\hwinjoneid              & \hwinjonefreq   & \hwinjonefonedot    & \hwinjonealpha     & \hwinjonedelta     & \hwinjonedfreq       & \hwinjonedfonedot          & \hwinjonedsky          \\
	\hwinjsevenid            & \hwinjsevenfreq & \hwinjsevenfonedot  & \hwinjsevenalpha   & \hwinjsevendelta   & \hwinjsevendfreq     & \hwinjsevendfonedot        & \hwinjsevendsky        \\
	\enddata
\end{deluxetable*}

Table \ref{tab:HIRecoveryMonteCarlo} lists the parameters of the hardware injections (transitioned to our reference time) and the offsets between these parameters and the posterior sample that maximized the $\fstat$-statistic in Stage 3.
The results showcase the high resolution of the signal space attainable with continuous gravitational waves.
The offsets are larger than those reported in \cite{Steltner:2023cfk} which were obtained using both LIGO O3a and O3b data, corresponding to an approximately twice as long time-baseline.
The uncertainties are within the expected range for an observation period of half a year, given the examined parameter-range.

\subsection{Upper Limits}\label{sec:upperlimits}

\begin{figure*}[ht]
	\includegraphics[width=\textwidth]{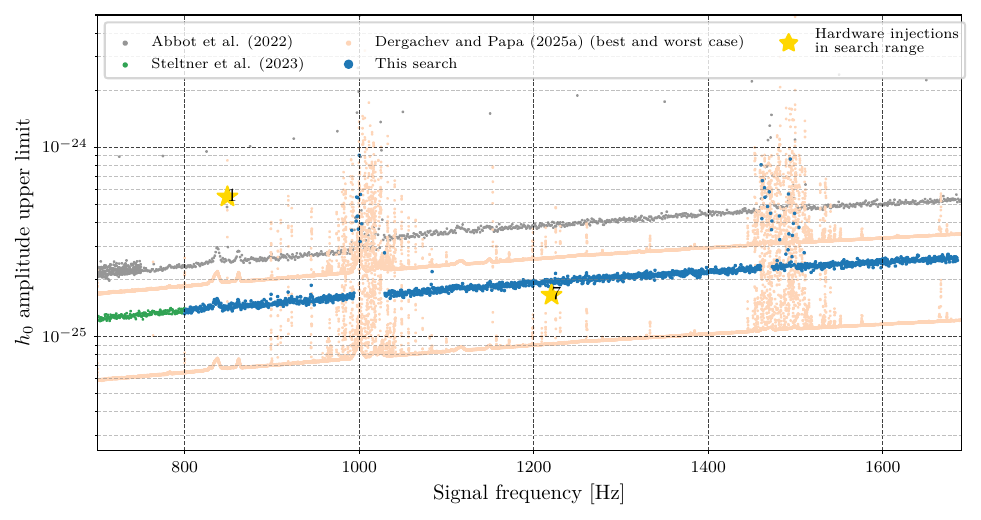}
	\caption{Smallest gravitational wave amplitude $h_0$ that our results exclude at 90\% confidence. We compare to other all-sky searches in LIGO O3 data \citep{KAGRA:2022dwb,Dergachev:2025hwp}. For \citet{Dergachev:2025hwp} two curves are shown, as they set best and worst case upper limits (circular and linear polarization). The golden stars mark the gravitational wave amplitude $h_0$ of the two hardware injections in our search range. We recover both. This marks the first successful recovery of hardware injection 7. }
	\label{fig:h0ULs}
\end{figure*}

We set upper limits on the gravitational wave amplitude $h_0$ in half-Hz bands using the same methodology as \citet{Steltner:2023cfk}.

As described in previous sections, the upper limit set in a half-Hz band may not hold in the whole band. The sub-bands excluded from the upper limit statements are given in the supplementary materials \citep{O3ASHi-AEI}. We recap here the different situations where this happens:

\begin{enumerate}
	\item the $\fiftyMHzband$ bands identified as disturbed by the visual inspection process (see Section \ref{subsec:Stage0Post}). This amounts to about $\approx 0.5\%$ of the searched frequency range.

	\item the $\fiftyMHzband$ bands where the number of seeds is lower than the average and associated with steep shoulders of partially-cleaned lines (end of Section \ref{subsec:Stage0Post}). These make up  $\approx 0.8\%$  of the searched frequency range.

	\item the 1 Hz region 1500-1501 Hz described in Section \ref{sec:BayesianFUResults}.

	\item the half-Hz bands that are so contaminated by spectral disturbances that 90\% detection efficiency cannot be achieved. This is not due to the contamination per se, but to our frequency-domain line-cleaning procedure of the input data that together with the line also removes any signal. There are 79 such bands.

\end{enumerate}

\begin{figure}[ht]
	\centering
	\includegraphics[width=\columnwidth]{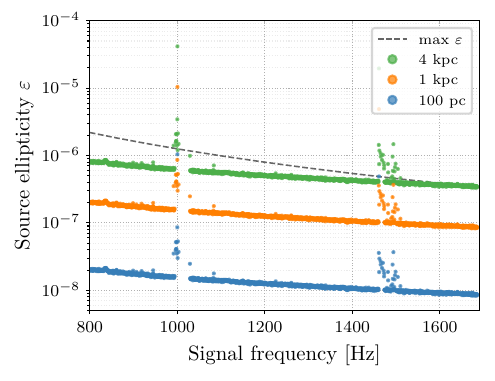}
	\caption{Upper limits on the neutron star ellipticity $\varepsilon$ at different distances. The dashed line shows the maximum ellipticity probed due to the maximum spin-down of this search.}
	\label{fig:epsilonULs}
\end{figure}

The upper limits on the gravitational wave amplitude $h_0$ can be translated to upper limits on the ellipticity $\varepsilon$ of a source modelled as a triaxial ellipsoid spinning around a principal moment of inertia axis ${I}$ at a distance $d$:

\begin{equation}
	\begin{split}
		\varepsilon^{\textrm{UL}} & =1.4 \times 10^{-6} ~\left( {h_0^{\textrm{UL}}\over{1.4\times10^{-25}}}\right ) \times                                                    \\
		                          & \left ( {d\over{1~\textrm{kpc}}}\right ) \left ({{\textrm{170~Hz}}\over f} \right )^2 \left ({10^{38}~{\textrm{kg m}}^2\over I} \right ). \\
	\end{split}
	\label{eq:epsilon}
\end{equation}
Figure \ref{fig:epsilonULs} shows the upper limits on the ellipticity $\varepsilon$ for different distances.

\begin{figure}[ht]
	\includegraphics[width=\columnwidth]{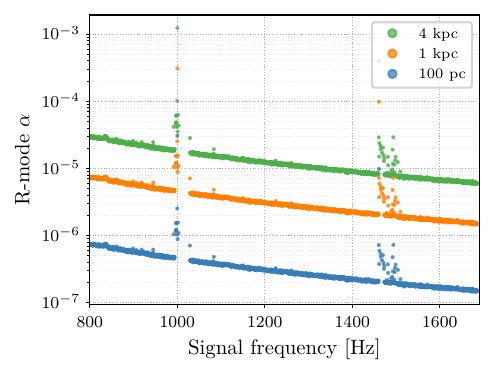}
	\caption{Upper limits on the r-mode amplitude $\alpha$ at different distances.}
	\label{fig:rmodeULs}
\end{figure}

Another possible emission mechanism are r-modes, unstable toroidal fluid oscillations driven by the Coriolis force, emitting at $\approx 4/3$ of the spin-frequency. We translate the upper limits on the gravitational wave amplitude $h_0$ in upper limits on the r-mode amplitude \citep{Owen:2010ng}:
\begin{equation}
	\alpha^{\textrm{UL}} = 0.028
	\left( \frac{h_0^{\textrm{UL}}}{\num{1e-24}} \right)
	\left( \frac{d}{1 \mathrm{kpc}} \right)
	\left( \frac{\SI{100}{\hertz}}{f} \right)^3.
\end{equation}
Figure \ref{fig:rmodeULs} shows the upper limits on the r-mode amplitude $\alpha$ for different distances $d$.

\section{Conclusions}\label{sec:Conclusions}

We present our all-sky search results for continuous gravitational waves with frequencies $\SI{\paramfmin}{\hertz} \leq f \leq \SI{\paramfmax}{\hertz}$ and spin-down $-2.7 \times 10^{-9} \text{Hz/s} \leq \dot{f} \leq 0.2 \times 10^{-9}$ Hz/s using LIGO O3 public data.
We identify the $\approx 13$ million most interesting continuous gravitational wave signal candidates from a total of $\paramtotaltemplates$ templates.
We recover all hardware injections within our parameter space, marking the first successful recovery of hardware injection 7.

Searching for such high frequency continuous gravitational wave signals allows one to probe interesting regimes of source ellipticity. Let us consider an extremely optimistic case, in order to determine the extent of ellipticitiy regimes we are possibly probing. For a neutron star as close as $10$ pc \citep{Pagliaro_2023} and spinning as the fastest observed pulsar, at $716\textrm{Hz}$ \citep{Hessels:2006ze}, assuming the canonical value for the sources moment of inertia of $10^{38}\textrm{kg m}^{2}$ our upper limits exclude ellipticities in the $10^{-10}$ region.
At the distance of the closest of the closest non-recycled neutron from \citep{Pagliaro_2023}, 1.9 kpc,  our results exclude ellipticities greater than $\approx 2\times 10^{-7}$ level.
These are interesting ranges \citep{Morales:2023euv, JohnsonMcDaniel:2012wg, Bhattacharyya:2020paf}.

This is the most sensitive all-sky search performed to date in this parameter space region. Our investigation of multiple search configurations makes us confident that this search setup achieves close to maximum sensitivity at the given computational cost.

This search uses a novel Bayesian method for the follow-up of signal candidates \citep{Martins:2025jnq}. Compared to past Einstein@Home searches, the new method significantly reduces the computational cost of the follow-up and the complexity of the search hierarchy.

\section*{Acknowledgments}
\begin{acknowledgments}
	We thank the E@H volunteers, without the support of whom this search could not have happened. We acknowledge the use of Topcat \citep{2005ASPC..347...29T}; Bilby \citep{Ashton:2018jfp} and scikit-learn \citep{pedregosa2011scikit}. This research has made use of data or software obtained from the Gravitational Wave Open Science Center (gwosc.org), a service of LIGO Laboratory, the LIGO Scientific Collaboration, the Virgo Collaboration, and KAGRA.
\end{acknowledgments}

\bibliography{references}

\end{document}